\documentclass[]{pasj01}
\Received{2020/03/19}%{yyyy/mm/dd}
\Accepted{2020/06/14}%{yyyy/mm/dd}
\usepackage{lscape}
\usepackage{graphicx}
\usepackage{bm}
\newcommand{\argmax}[1]{\mathop{\rm arg~max}_{#1}\limits}

\begin{document} 

\title{Feature Selection for Classification of Blazars Based on Optical Photometric and Polarimetric Time-Series Data}

%%% begin:list of authors
% Do NOT capitalize all letters in "textsc".
\author{Makoto \textsc{Uemura}\altaffilmark{1}}%
\altaffiltext{1}{Hiroshima Astrophysical Science Center, Hiroshima
  University, 1-3-1 Kagamiyama, Higashi-Hiroshima, 739-8526, Japan}
\email{uemuram@hiroshima-u.ac.jp}

\author{Taisei \textsc{Abe}\altaffilmark{1}}

\author{Yurika \textsc{Yamada}\altaffilmark{1}}

\author{Shiro \textsc{Ikeda}\altaffilmark{2}}
\altaffiltext{2}{The Institute of Statistical Mathematics, 10-3
  Midori-cho, Tachikawa, Tokyo 190-8562, Japan}
%%% end:list of authors

%% `\KeyWords{}' always has to be placed before ``\maketitle'' 
%%  List of Key Words:  https://academic.oup.com/pasj/pages/Pasj_Keywords 
\KeyWords{BL Lacertae objects: general --- galaxies: active --- galaxies: statistics}

\maketitle

\begin{abstract}
Blazars can be divided into two subtypes, flat spectrum radio quasars
(FSRQs) and BL~Lac objects, which have been distinguished
phenomenologically by the strength of their optical emission lines,
while their physical nature and relationship are still not fully
understood. In this paper, we focus on the differences in their
variability. We characterize the blazar variability using the
Ornstein--Uhlenbeck (OU) process, and investigate the features that
are discriminative for the two subtypes. We used optical photometric
and polarimetric data obtained with the 1.5-m Kanata telescope for
2008--2014. We found that four features, namely the variation
amplitude, characteristic timescale, and non-stationarity of the
variability obtained from the light curves and the median of the
degree of polarization (PD), are essential for distinguishing between
FSRQs and BL~Lac objects. FSRQs are characterized by rare and large
flares, while the variability of BL~Lac objects can be reproduced with
a stationary OU process with relatively small amplitudes. The
characteristics of the variability are governed
not by the differences in the jet structure between the subtypes, but
by the peak frequency of the synchrotron emission. This implies that
the nature of the variation in the jets is common in FSRQs and BL~Lac
objects. We found that BL~Lac objects tend to have high PD medians,
which suggests that they have a stable polarization component. FSRQs
have no such component, possibly because of a strong Compton cooling
effect in sub-pc scale jets. 
\end{abstract}

\section{Introduction}

Blazars are a sub-class of active galactic nuclei (AGN) with
relativistic jets that point toward us. The jet emission is
amplified by the beaming effect and dominates the observed flux at
almost all wavelengths
(\cite{bla78bllac,bla79agn,urr95agn}). Synchrotron emission from the
jet is dominant in the radio--X-ray regime. The observed
X-ray--$\gamma$-ray emission is mostly due to inverse-Compton
scattering by relativistic electrons in the jet. Blazars exhibit 
violent variability, which provides a hint for understanding the
physical conditions and structure in AGN jets (e.g. \cite{ulr97var}).

Blazars consist of two subtypes: flat spectrum radio quasars (FSRQs)
and BL Lac type objects. The former was originally defined by strong
emission lines observed in the optical spectra
(equivalent width $> 5$~\AA; \cite{sti91bllac,sto91bllac}), while the
latter was defined by weaker lines or featureless spectra. In
addition, FSRQs have lower peak frequencies in the synchrotron
emission, $\nu_{\rm peak}\lesssim 10^{14}\;{\rm Hz}$, in their spectral
energy distribution (SED), while BL~Lac objects have a
wide range of $\nu_{\rm peak}$
($10^{14}\lesssim \nu_{\rm peak}\;[{\rm Hz}]\lesssim 10^{18}$)
(\cite{abd10sed}). The luminosity of blazars has a negative
correlation with $\nu_{\rm peak}$; FSRQs form the most luminous class
of blazars, while BL~Lac objects are less luminous. In SEDs, the
relative strength of the inverse-Compton scattering component to the
synchrotron component is larger in FSRQs than in BL~Lac objects. These 
regularities are known as the ``blazar sequence''
(\cite{ghi98seq}).

In addition to the classification based on the emission line strength,
a classification scheme based on $\nu_{\rm peak}$ is also used for
blazars, with low synchrotron peaked (LSP) blazars for objects with
$\nu_{\rm peak}\lesssim 10^{14}\;{\rm Hz}$, intermediate synchrotron
peaked (ISP) blazars with 
$10^{14}\lesssim\nu_{\rm peak}\;[{\rm Hz}]\lesssim 10^{15}\;{\rm Hz}$,
and high synchrotron peaked (HSP) blazars with
$\nu_{\rm peak}\gtrsim 10^{15}\;{\rm Hz}$ (\cite{abd10sed}).
Most FSRQs are LSP blazars. In this paper, we call LSP, ISP, and
HSP BL~Lac objects LBLs, IBLs, and HBLs, respectively.

The nature and links between blazar subtypes are still incompletely
understood. \citet{ghi08seq} proposed that FSRQs are AGN
having a radiatively efficient accretion disk (a ``standard'' disk; 
\cite{sha73disk}) with a high accretion rate, while BL~Lac objects
have a radiatively inefficient accretion flow (RIAF;
\cite{qua01riaf,nar95adaf}) with a low accretion rate. The accretion
rate is considered to be linked to the extended radio morphology of
radio galaxies, that is, the Fanaroff--Riley (FR) classification
(e.g., \cite{bau95fr}). It is proposed that FSRQs and all or some LBLs 
are beamed counterparts to FR type II radio galaxies with high
luminosity, and IBLs, HBLs, and possibly some LBLs are counterparts to FR
type I objects with low luminosity (\cite{mey11type,gio12type}).
\citet{gio12type} report that known LBLs are inhomogeneous and contain
both FR~I and II subtypes. 

The variability characteristics of the flux and polarization have also
been discussed for the different subtypes of blazars, particularly for
the optical waveband in which all subtypes have been frequently
monitored. It is well known that the optical activity apparently
depends on $\nu_{\rm peak}$; LSP blazars are more variable than HSP
blazars (e.g., \cite{bau09var,ike11blazar,hov14var}). A similar
$\nu_{\rm peak}$ dependence has also been reported in the polarization
variations, though the number of previous studies is limited
(\cite{ito16fermi,ang16pol}). The mechanism of the effect of the
$\nu_{\rm peak}$ on the observed flux and polarization variability is
unclear. High $\nu_{\rm peak}$ objects show less activity, possibly
because a large number or large area of emitting regions blur each
short flare (\cite{mar10multi,ito16fermi,ang16pol}), or possibly
because the jet volume fraction of a slower ``sheath'' component
increases (\cite{ito16fermi,ghi05spine}). 

In this paper, we focus on blazar variability. We have performed
photometric and polarimetric monitoring of blazars using the 1.5-m
Kanata telescope in Hiroshima since 2008 
(\cite{ike11blazar,ito16fermi}). The present study has two major
objectives: to establish the observational features of the flux and
polarization variability for characterizing the subtypes, and thereby
to investigate the nature of the subtypes, for example, whether FSRQs
and LBLs have a common origin and whether the jet structure of FSRQs
is different from that of BL Lac objects. Our analyses can be divided
into two parts, the extraction of features from the observed
time-series data and the selection of the features which are
discriminative for the two subtypes. For the feature extraction, 
in past studies the blazar variability was occasionally characterized
only by the features based on the variance of the whole data,
while the variation timescale was not considered. We use the
Ornstein--Uhlenbeck (OU) process to estimate both the timescale and
the amplitude from the data. The OU process and more advanced models
based on it have been used to characterize the variations observed in
AGN and also in blazars
(\cite{kel09agn,kel11mixou,rua12ou,sob14ou}). For the feature 
selection, we propose a data-driven approach to select the best set of 
features for classifying blazars by maximizing the
generalization error of a classifier. 

The structure of this paper is as follows: In \S~2, we describe
the data (\S~2.1) and methods used in this paper, namely, the OU
process for the feature extraction (\S~2.2) and sparse multinomial
logistic regression for the classifier (\S~2.3). In \S~3, we present
the results of the feature selection. In \S~4, we evaluate the
classifier and discuss the implications for the selected features. 

\section{Data and methods}

\subsection{Data}

\begin{figure*}
 \begin{center}
  \includegraphics[angle=90,width=17cm]{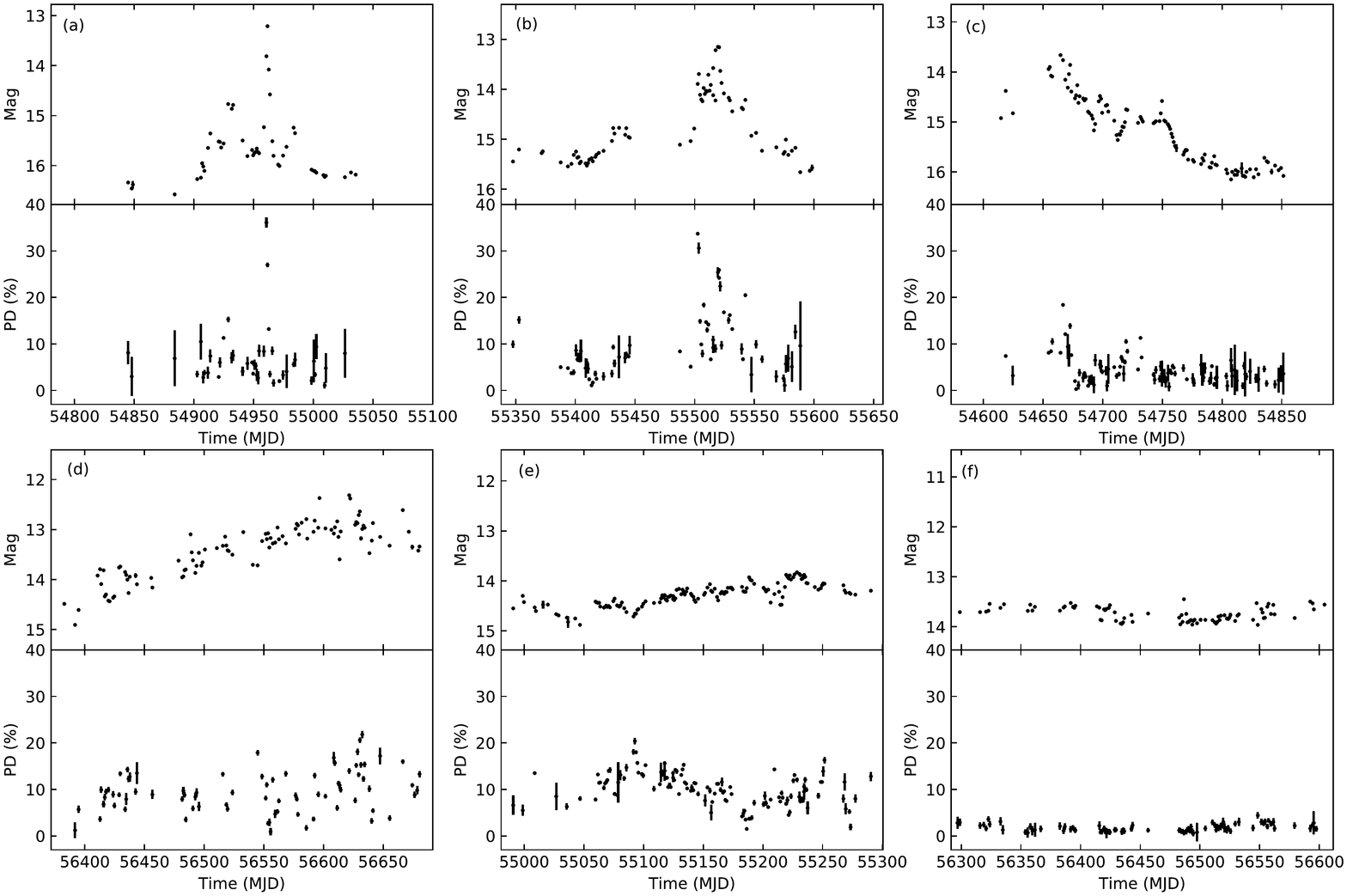} 
 \end{center}
 \caption{Examples of light curves and PD time-series data used in
 this paper. (a) PKS~1510$-$089 (FSRQ), (b) and (c) 3C~454.3 (FSRQ) in
 2010 and 2008, respectively, (d) BL~Lac (LBL) in 2013, (e) 3C~66A
 (IBL) in 2009, and (f) Mrk 501 (HBL). The upper and lower panels show
 the light curves and PD variations for each, respectively. The
 vertical and horizontal scales are common in all
 panels.}\label{fig:samples}
\end{figure*}

We used the data obtained with the Kanata telescope which was published in
\citet{ito16fermi}. The data includes $V$-band time-series photometric
and polarimetric data for 45 blazars from 2008--2014.
Figure~\ref{fig:samples} shows examples of light curves and variations
in the degree of polarization (PD).

Panels~(a), (b), and (c) of figure~\ref{fig:samples} show examples of
FSRQs: PKS~1510$-$089 and 3C~454.3 in 2010 and in 2008, respectively.
The peak-to-peak amplitudes in the light curves are large, over 2~mag
in all cases, while the light curve profiles are diverse: a solitary,
short flare appears in panel~(a), while a number of short flares 
superimposed on long outbursts appear in panels~(b) and (c). The light
curves change their apparent characteristics year by year even for the
same object, as shown in panels~(b) and (c). Panels~(d), (e), and (f)
of figure~\ref{fig:samples} show examples of BL Lac objects: BL~Lac
(LBL), 3C~66A (IBL), and Mrk~501 (HBL), respectively. The peak-to-peak 
amplitude of the light curve in panel~(d) is comparable to that of
FSRQs, while the light curve profile looks different. The
characterization and classification of these variations are the 
main subjects of this paper. The variation in polarization could
give rise to some interesting features. For example, PD flares are
associated with FSRQ flares, though no clear correlation can be seen
in the light curve and PD variations in panels~(d) and (e). Panels~(e)
and (f) show that the variation amplitude apparently decreases from
LBL to HBL, as mentioned in the previous section. 

\subsection{Feature extraction with the OU process}

We use the OU process for our time-series analysis. The OU process is
a stochastic model based on the multivariate normal distribution whose
covariance between the data at time $t_i$ and $t_j$, $S_{ij}$, is
given as: 
\begin{eqnarray}
  S_{ij}=A_{\rm exp}\exp\left(-\frac{|t_i-t_j|}{\tau}\right),
\end{eqnarray}
where $\tau$ and $A_{\rm exp}$ represent the characteristic variation
timescale and amplitude at $\tau$, respectively (\cite{ou30}). For
the time-series data followed by the OU process, $f(t)$, the 
observed data, $m(t)$, is given by
$m(t)=f(t)+\mathcal{N}(0,\sigma_{\rm OU}^2)$, where the second term is
the noise defined by the normal distribution having zero mean and
variance $\sigma_{\rm OU}^2$. We can extract the characteristic
features, $A_{\rm exp}$, $\tau$, and $\sigma_{\rm OU}^2$, from the
observed time-series data using the OU process regression.

The time-series data introduced in \S~2.1 have different observation
periods for each object. The time-series data of each object was
divided into one-year segments, each of which is regarded as a sample
in this paper. For modeling the light curves with the OU process, the
magnitude values were translated to fluxes on a logarithmic scale,
simply dividing by $-2.5$. We assumes that the light curves
are approximated with the OU process with a characteristic time-scale
less than a few tens of days for our sample. The short time-scale is
supported by the data in which erratic variations are detected over
measurement errors in all samples. If our assumption is true, the
power spectrum should be flat for frequencies ($f$) lower than the
characteristic frequency, and decays as $f^{-2}$ for higher
frequencies (\cite{kel11mixou}). A strong linear trend in the
time-series data breaks this assumption because the power becomes
larger in lower frequencies. We consider that the linear trend has an
origin different from the short-term variations governed by the OU
process. The presence of such distinct short- and long-term variations
are reported in AGN (\cite{Arevalo2006, McHardy2007, kel11mixou}) and
also in blazars (\cite{sob14ou}). Hence, we first subtracted the linear
trend from the samples, and then performed the OU process regression.
The slope value of the linear trend can be considered an indicator of 
the power at the lowest frequencies, and we use it as a feature for
the classification in the next section.  

\begin{figure}
 \begin{center}
  \includegraphics[width=8cm]{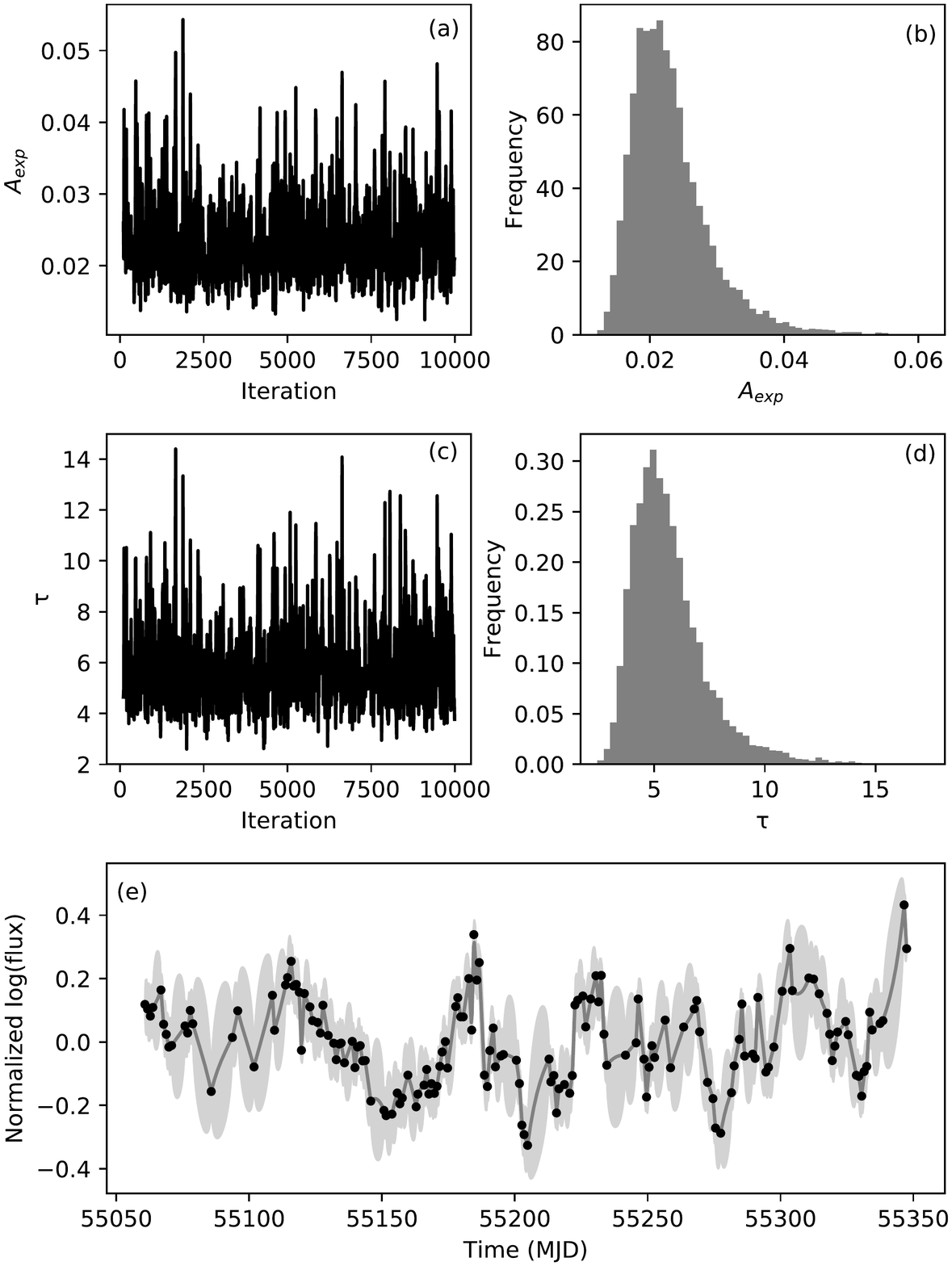} 
 \end{center}
 \caption{Results of MCMC estimation of $A_{\rm exp}$ and $\tau$
   for the sample S5~0716$+$714 between MJD~55050 and
   55389. Panels~(a) and (c) are the trace plots of the MCMC samples
   of $A_{\rm exp}$ and $\tau$, respectively. Panels~(b) and (d) are
   their posterior probability distributions. Panel~(e) is the
   observed and model light curves. The filled circles are the
   data and the solid line and shaded region indicate the mean of the
   model prediction and its 95\% confidence interval,
   respectively. This is an example in which both $A_{\rm exp}$ and
   $\tau$ are uniquely determined.}\label{fig:good}
\end{figure}

The OU process is identical to the Gaussian process with an
exponential kernel. We used the {\tt python} package for the Gaussian
process, {\tt GPy}, which includes a package for the Markov chain Monte
Carlo (MCMC) method for the estimation of the posterior probability
distributions of the parameters. In the present work, we estimated the
posterior distribution of $A_{\rm exp}$ with a flat prior probability
and that of $\tau$ with a positive flat prior. We fixed
$\sigma_{\rm OU}^2$ with a typical measurement error of the data. 
We estimated the posterior probability distributions of $A_{\rm exp}$
and $\tau$ using the MCMC method for each light-curve sample. We set
$\sigma_{\rm OU}^2=10^{-5}$. Figure~\ref{fig:good} shows trace plots
of $A_{\rm exp}$ and $\tau$, their posterior distributions, and the
observed and model light curves for the sample S5~0716$+$714 between
MJD~55050 and 55389. The MCMC samples converge to a stationary
distribution and the posterior distributions have single-peaked
profiles. In this case, we successfully obtained unique solutions of
$A_{\rm exp}$ and $\tau$. 

\begin{figure}
 \begin{center}
  \includegraphics[width=8cm]{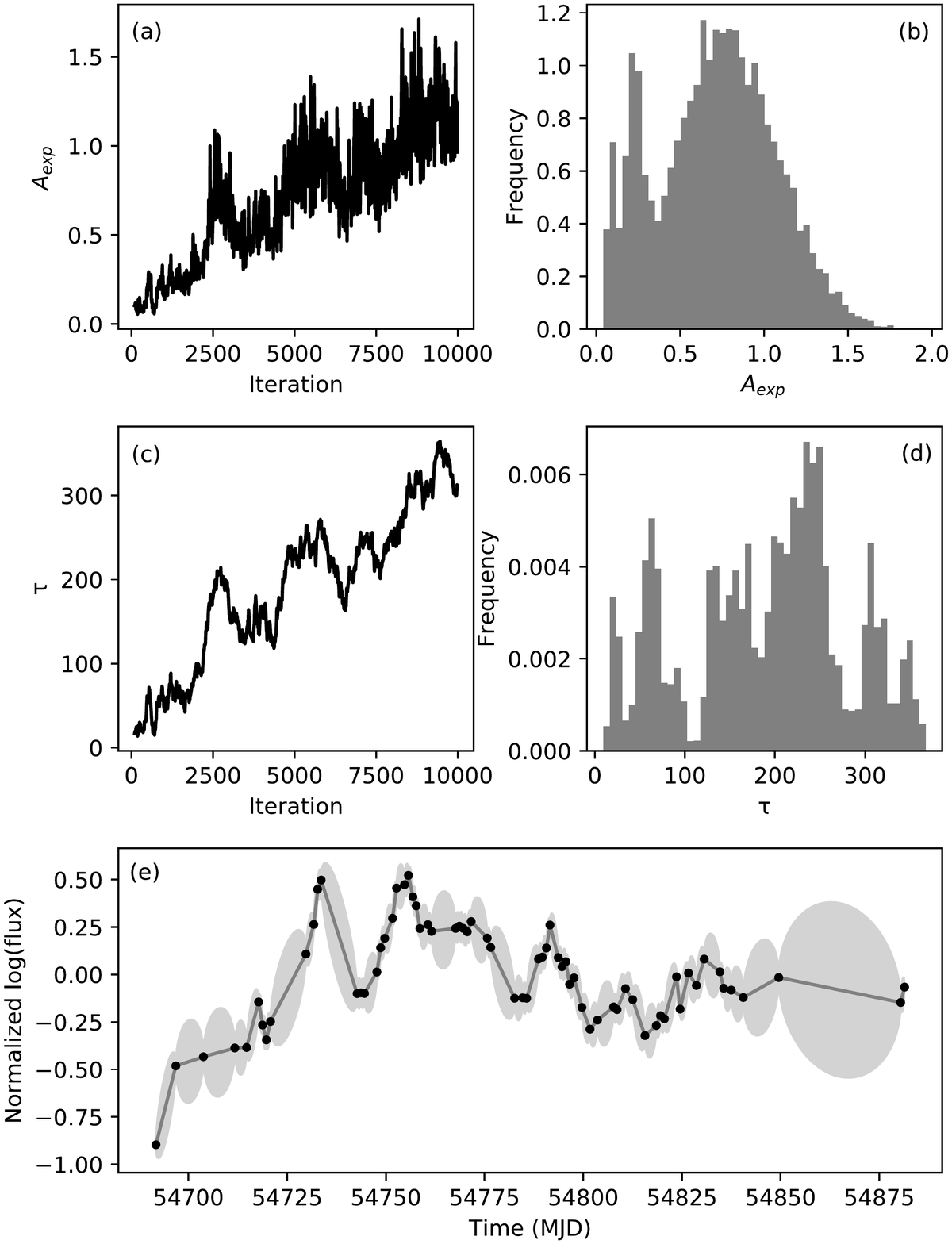} 
 \end{center}
\caption{As for figure~\ref{fig:good} with the light curve
  data of AO~0235$+$16 between MJD~54617 and 54946. This is an example
  in which $\tau$ is too long.}\label{fig:bad}
\end{figure}

On the other hand, we found that $A_{\rm exp}$ and $\tau$ were not
uniquely determined for several of the samples, mainly because the
data size is not large enough to make a meaningful estimate of the
parameters. A significant number of the samples from \citet{ito16fermi}
have only $<30$ data points. Even in the samples with larger data
size, $\tau$ is not uniquely determined if it is too long.
Figure~\ref{fig:bad} shows an example, AO~0235$+$16 between MJD~54617
and 54946. The MCMC samples do not converge to a stationary
distribution and $\tau$ can be very large, reaching over
300~d. \citet{koz17drw} reports that the OU process model is
degenerate when the baseline of the light-curve sample is shorter
than ten times $\tau$. The result in figure~\ref{fig:bad} is probably an
example of such a case.. 

In this paper, we used only samples for which $A_{\rm exp}$ and
$\tau$ were uniquely determined, as in figure~\ref{fig:good}. This
selection reduces the number of samples to 38 for 18 objects. The
selected samples include 12 samples for 8 FSRQs and 26 samples for 10
BL~Lac objects. The samples are listed in table~\ref{tab:log} in the
Appendix. The designation of the objects to the subtypes FSRQ, LBL,
IBL, and HBL is taken from \citet{ito16fermi}. 

Blazars occasionally exhibit large prominent flares, as shown in
panel~(a) of figure~\ref{fig:samples}, which definitely arise from a
non-stationary process, whereas the OU process model assumes a
stationary stochastic process. In order to characterize the
non-stationarity, we calculate the cross-validation error (CVE) using
the OU process regression, as follows: First, the sample is divided
into 25-d bins, a sub-sample of which is for validation while the
others are for training. Then, the OU process regression is performed
with the training subsets. Then, the log-likelihood is calculated from
the validation subset and the optimized model. Using the other
sub-samples as validation data, we obtained about $10$ log-likelihoods
for each sample. The CVE is defined as their mean. A large CVE
means that the validation data has a large deviation from the
prediction of the model constructed from the training data. Hence, a
large value of CVE indicates a high degree of non-stationarity.

\begin{figure}
 \begin{center}
  \includegraphics[width=8cm]{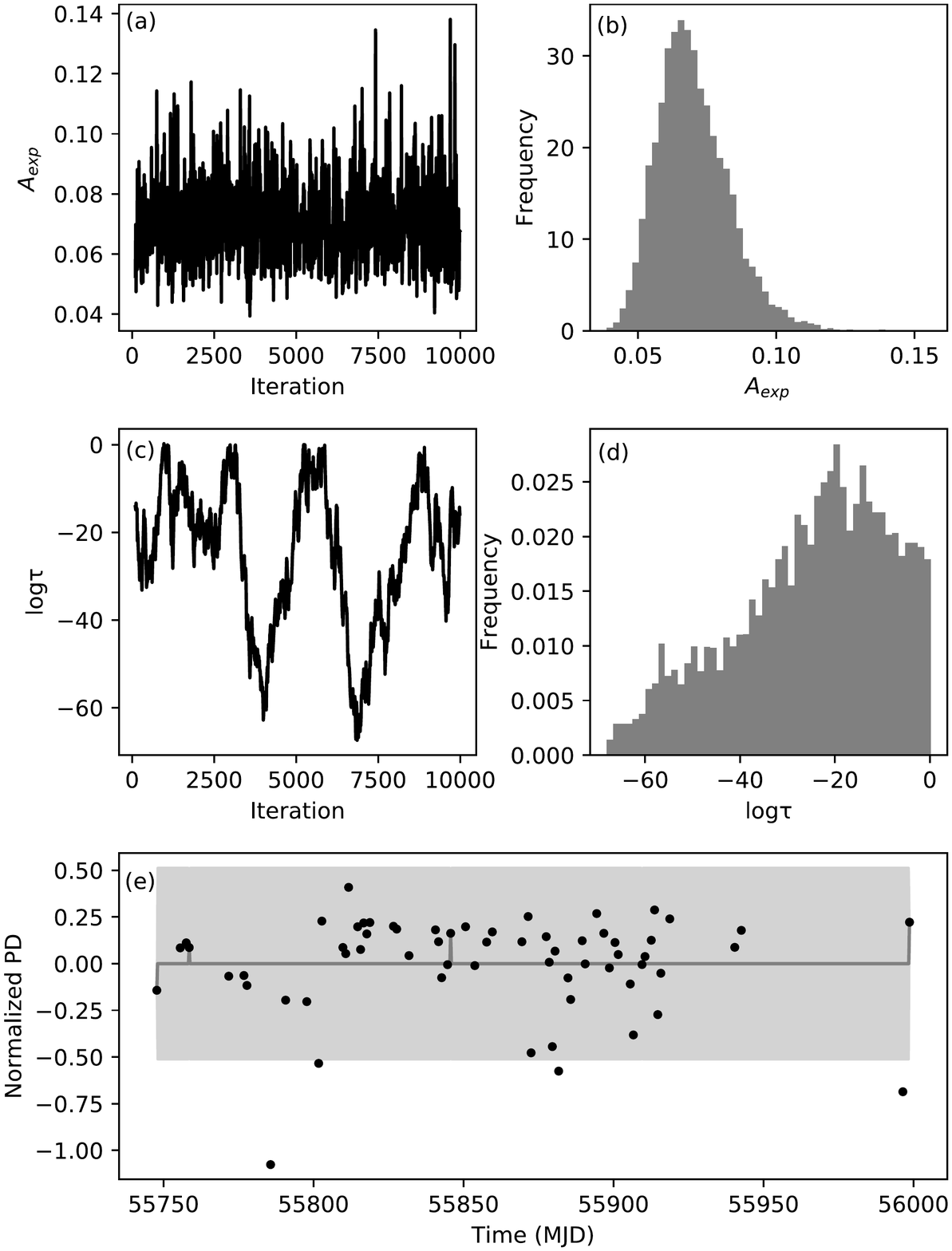} 
 \end{center}
\caption{As for figure~\ref{fig:good} with the PD data of
  S5~0716$+$714 between MJD~55746 and 56125. Note that the scale of
  $\tau$ is logarithmic. This is an example of PD analysis in
  which $\tau$ is too short.}\label{fig:pdbad}
\end{figure}

The analysis of the time-series PD data was performed in the
same manner as for the light curves, that is, dividing it into
one-year-segments, converting to a logarithmic scale whose linear
trends were subtracted. The OU process parameter
$\sigma_{\rm OU}^2$ was fixed to $10^{-4}$. $A_{\rm exp}$ and $\tau$ were
uniquely determined for the PD variations in all the samples except for
seven. An example of one of the seven samples is shown in
figure~\ref{fig:pdbad}. Although, $A_{\rm exp}$ is uniquely
determined, the MCMC samples of $\tau$ do not converge, and $\tau$ can
be quite small (note that the scale of $\tau$ is logarithmic in
figure~\ref{fig:pdbad}). As a result, the model of PD is simply the mean
of the data, as shown in panel~(e). These results suggest that the
characteristic timescale is too short to be properly determined with
our data. We set $\tau=0.0$ for the seven samples. While a value of
zero for $\tau$ is physically undefined, it 
works for training and evaluating the classifier representing very
short timescales. CVEs were also calculated for the PD data. In
addition, we also used the median of the PD as a feature parameter. 

In total, we obtained nine features from the light curve and
PD data: the four features from the light curve samples,
$A_{\rm exp}$, $\tau$, the slope of the linear trend, and CVE, and
five features from the PD samples, $A_{\rm exp}$, $\tau$, linear
slope, CVE, and PD median. The values of the features are listed in
table~\ref{tab:log} in the Appendix.

\subsection{Sparse multinomial logistic regression}

We construct a classifier for FSRQs and BL~Lac objects based on the
nine features described in \S~2.2. We use sparse multinomial
logistic regression (SMLR) to determine the classifier
(\cite{smlr}). We consider the problem of defining an $M$-class
classifier with $N$ labeled samples, each of which has a
$K$-dimensional feature vector, 
$\bm{\theta}_i=\{\theta_{i,1},\theta_{i,2},\cdots,\theta_{i,K}\}$
$(i=1,2,\cdots,N)$. A sample that belongs to the $j$-th class can be
expressed with a vector 
$\bm{y}=\{ y^{(1)}, y^{(2)}, \cdots , y^{(M)} \}$ such that
$y^{(j)}=1$ and the other elements are $0$. Multinomial logistic
regression gives the probability that a sample belongs to the $j$-th
class, as follows:
\begin{eqnarray}
  P\left(y^{(j)}=1|\bm{\theta},\bm{w}\right)
  = \frac{\exp\left( {\bm{w}^{(j)}}^T \bm{\theta} \right)}
  {\sum_{j=1}^M \exp\left( {\bm{w}^{(j)}}^T \bm{\theta} \right)},
\end{eqnarray}
where $\bm{w}^{(j)}$ is the weight vector for the $j$-th class. The
log-likelihood function is given by the data $\bm{\theta}$ as
\begin{eqnarray}
  \ell(\bm{w})=\sum_{i=1}^N \log P\left(\bm{y}_i|\bm{\theta}_i,\bm{w}\right).
\end{eqnarray}
Then, the solution of SMLR is expressed as
\begin{eqnarray}
  \hat{\bm{w}}= \argmax{\bm{w}} \{ \ell(\bm{w}) - \lambda\|\bm{w}\|_1 \},
\end{eqnarray}
where $\|\bm{w}\|_1$ is the $\ell_1$ norm, $\|\bm{w}\|_1=\sum_i|w_i|$,
and $\lambda$ is a sparsity parameter that controls the complexity of
the model.

SMLR gives a linear classifier against the observed features if it
is used as $\bm{\theta}$. In this case, SMLR can select the important
features because the $\ell_1$ term makes $\bm{w}$ sparse. On the other
hand, a non-linear classifier can be obtained if the observed features
are transformed with non-linear kernel functions. Then, we can avoid
over-fitting due to the $\ell_1$ term. In the present study, the
features listed in table~\ref{tab:log} were normalized and the feature
vector of the $i$-th sample, $\bm{x}_i$, was obtained. The $j$-th
element of $\bm{\theta}_i$ was obtained from $\bm{x}_i$ and $\bm{x}_j$
with the RBF kernel as follows:
\begin{eqnarray}
  \theta_{i,j} = \exp\left\{ -\frac{|\bm{x}_j-\bm{x}_i|^2}{2\sigma_{\rm RBF}^2} \right\}, 
\end{eqnarray}
where $\sigma_{\rm RBF}^2$ is the bandwidth.

As mentioned in \S~2.2, the number of samples, $N$, is 38. The
number of classes, $M$, is two: FSRQs and BL~Lac objects. Because of
the small sample size, the three subtypes LBL, IBL, and HBL are
combined as one BL~Lac type, while the characteristics of the subtypes
are discussed in \S~4.2. The classifier is evaluated from the
so-called ``area under the curve'' (AUC), which is defined by the
receiver operating characteristic (ROC) curve. The simple accuracy of
the classifier is inadequate because the number of BL~Lac objects is
larger than that of FSRQs in our sample (12 FSRQ samples and 26
BL~Lac samples). The AUC is calculated by leave-one-out
cross-validation for estimating the generalization error of the
classifier. Optimization of the model and the calculation of
the cross-validated AUC were performed with the Java-based application
{\tt SMLR}\footnote{http://www.cs.duke.edu/\~{}amink/software/smlr}.

\section{Results}

We investigate the features that are discriminative for FSRQs and
BL~Lac objects based on SMLR and cross-validated AUC using the nine
features obtained from the data. SMLR has two hyper-parameters,
$\sigma_{\rm RBF}^2$ and $\lambda$. We first consider appropriate 
values for these two parameters for our study. 

\begin{figure}
 \begin{center}
  \includegraphics[width=8cm]{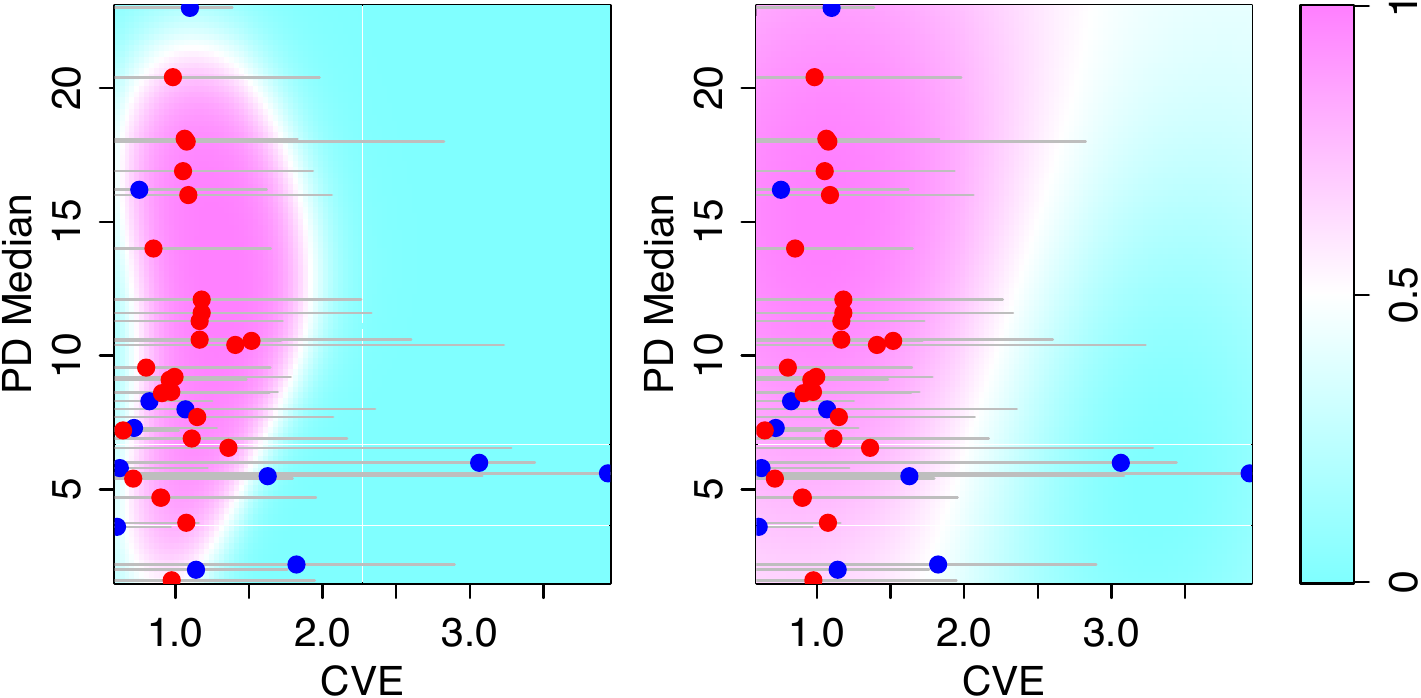} 
 \end{center}
 \caption{Examples of complicated and simple boundaries. The color
   map indicates the probability map of a BL~Lac type sample calculated
   from the light-curve CVE and PD median with SMLR. The left and
   right panels show those obtained with bandwidth parameters of
   1.0 and 5.0, respectively. The blue and red circles indicate FSRQ
   and BL~Lac samples.}\label{fig:island} 
\end{figure}

A small $\sigma_{\rm RBF}^2$ leads to a complicated model with a
large number of samples retained in the classifier. Such a small
$\sigma_{\rm RBF}^2$ occasionally creates an island-like boundary.
Figure~\ref{fig:island} shows examples of the probability map of
BL~Lac type samples calculated with $\sigma_{\rm RBF}^2=1.0$ (left)
and $5.0$ (right) using two features, the light-curve CVE and PD
median. We set $\lambda=0.1$ in this case. As can be seen in the left
panel, the high probability region forms an ``island'' within the
surrounding low probability area. However, it is unlikely that the two
subtypes of blazars have such a complicated boundary. A linear or
slightly non-linear model, like that in the right panel, is more
reasonable. We confirmed that a classifier with large bandwidths
($\sigma_{\rm RBF}^2\gtrsim 5$) does not have an island-like boundary 
using our samples. In the following analysis, we set 
$\sigma_{\rm RBF}^2=5.0$. 

\begin{figure}
 \begin{center}
  \includegraphics[width=8cm]{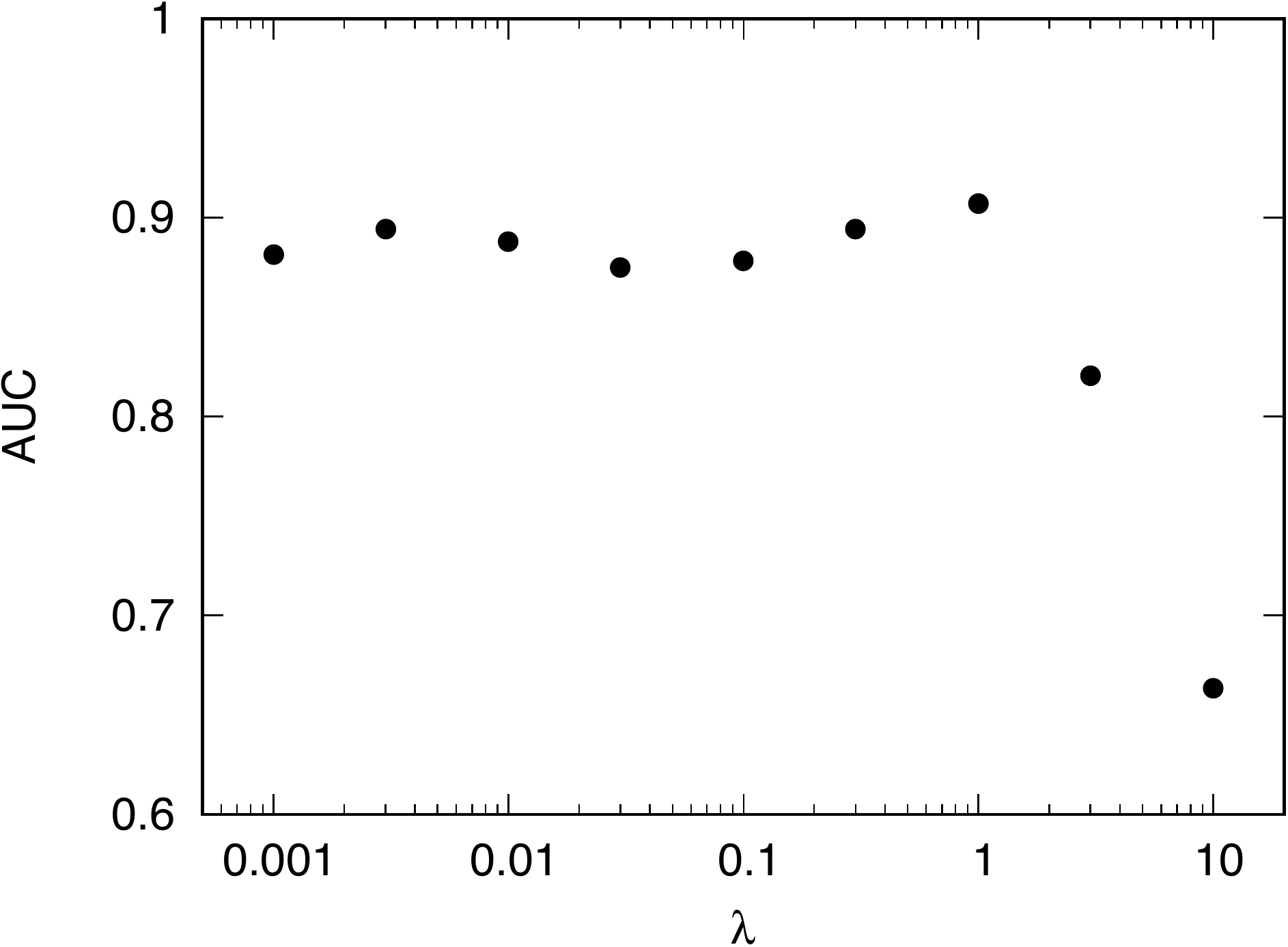} 
 \end{center}
 \caption{Optimal AUC against $\lambda$.}\label{fig:lambda} 
\end{figure}

The sparsity parameter, $\lambda$, also controls the complexity of the
model. We investigated the best AUC of all combinations of the nine
features against various values of $\lambda$. The result is given in
figure~\ref{fig:lambda}, showing that AUC becomes maximum around
$\lambda=1.0$. A model obtained with $\lambda>1$ is too simple to
appropriately classify the samples. On the other hand, a small
$\lambda$ ($<1$) leads to over-fitting. We set $\lambda=1.0$ in the
following analysis.

\begin{table*}
  % all01.dat
  \tbl{Variables used, AUC, and accuracy of top 20 models. A bullet symbol indicates the parameter was included in that model.}{%
    \begin{tabular}{ccccccccccrr}
      \hline
      \multicolumn{4}{c}{Light curve} &
      &\multicolumn{5}{c}{Polarization degree} & & \\
      \cline{1-4} \cline{6-10}
      CVE & Slope & $A_{\rm exp}$ & $\tau$ & & Median & CVE & Slope &
      $A_{\rm exp}$ & $\tau$ & AUC & Accuracy\\
      \hline
      $\bullet$ & --- &$\bullet$ & $\bullet$ & &
      $\bullet$ & --- &$\bullet$ & --- & --- & 0.907 & 0.842\\
      $\bullet$ & $\bullet$ &$\bullet$ & $\bullet$ & &
      $\bullet$ & --- & --- & --- & --- & 0.904 & 0.816\\
      $\bullet$ & $\bullet$ & $\bullet$ & $\bullet$ & &
      $\bullet$ & --- & $\bullet$ & --- & --- & 0.888 & 0.842\\
      $\bullet$ & $\bullet$ &$\bullet$ & $\bullet$ & &
      $\bullet$ & --- & --- & --- & $\bullet$ & 0.885 & 0.789\\
      $\bullet$ & $\bullet$ &$\bullet$ & $\bullet$ & &
      $\bullet$ & $\bullet$ & --- & --- & --- & 0.881 & 0.868\\
      $\bullet$ & --- & $\bullet$ & $\bullet$ & &
      $\bullet$ & --- & $\bullet$ & --- & $\bullet$ & 0.881 & 0.737\\
      $\bullet$ & $\bullet$ & $\bullet$ & $\bullet$ & &
      $\bullet$ & $\bullet$ & $\bullet$ & $\bullet$ & $\bullet$ & 0.878 & 0.816\\
      $\bullet$ & --- &$\bullet$ & $\bullet$ & &
      $\bullet$ & $\bullet$ & $\bullet$ & --- & --- & 0.878 & 0.789\\
      $\bullet$ & --- & $\bullet$ & $\bullet$ & &
      $\bullet$ & $\bullet$ & --- & --- & $\bullet$ & 0.875 & 0.789\\
      $\bullet$ & $\bullet$ &$\bullet$ & $\bullet$ & &
      $\bullet$ & $\bullet$ & --- & $\bullet$ & $\bullet$ & 0.872 & 0.789\\
      $\bullet$ & $\bullet$ &$\bullet$ & $\bullet$ & &
      $\bullet$ & $\bullet$ & --- & --- & $\bullet$ & 0.872 & 0.816\\
      $\bullet$ & $\bullet$ &$\bullet$ & $\bullet$ & &
      $\bullet$ & $\bullet$ & $\bullet$ & --- & --- & 0.869 & 0.842\\
      $\bullet$ & --- & $\bullet$ & $\bullet$ & &
      --- & $\bullet$ & --- & --- & --- & 0.865 & 0.842\\
      $\bullet$ & $\bullet$ & $\bullet$ & $\bullet$ & &
      $\bullet$ & $\bullet$ &$\bullet$ & $\bullet$ & --- & 0.862 & 0.763\\
      $\bullet$ & $\bullet$ & $\bullet$ & $\bullet$ & &
      $\bullet$ & --- & $\bullet$ & --- & $\bullet$ & 0.862 & 0.737\\
      $\bullet$ & $\bullet$ &$\bullet$ & --- & &
      $\bullet$ & --- & $\bullet$ & --- &$\bullet$ & 0.859 & 0.789\\
      $\bullet$ & $\bullet$ & $\bullet$ & $\bullet$ & &
      $\bullet$ & --- & $\bullet$ & $\bullet$ & $\bullet$ & 0.856 & 0.763\\
      $\bullet$ & --- & $\bullet$ & $\bullet$ & &
      $\bullet$ & $\bullet$ & $\bullet$ & $\bullet$ & --- & 0.856 & 0.763\\
      $\bullet$ & $\bullet$ & $\bullet$ & $\bullet$ & &
      $\bullet$ & $\bullet$ & $\bullet$ & --- & $\bullet$ & 0.853 & 0.816\\
      $\bullet$ & $\bullet$ & $\bullet$ & $\bullet$ & &
      $\bullet$ & $\bullet$ & --- & $\bullet$ & --- & 0.853 & 0.789\\
      \hline
\end{tabular}}\label{tab:auc}
\end{table*}

We made an exhaustive test of all the parameters to find the most
important features (e.g., \cite{iga18ex}). The number of combinations
of the nine features is $2^9-1=511$. Using SMLR, we developed 511
classifiers using models with different combinations of parameters,
and calculated the AUCs for each. Table~\ref{tab:auc} lists the top 20
classifiers in the order of AUC values. For example, the classifier
with the highest AUC ($=0.923$) uses six features, that is, the CVE,
$A_{\rm exp}$, $\tau$ of the light curve, and the median, CVE, and
$\tau$ of the PD. It has an accuracy of 0.842. As can be seen in the
table, the correlation between the accuracy score and AUC is low in
the 20 models. This is probably due to the small sample size, and
indicates that a small difference in the AUC is not
important. We found that CVE and $A_{\rm exp}$ of the light
curve are used in all the top 20 models, and that $\tau$ of the light
curve and PD median are used in 19 models.  This result
suggests that these four features are essential to classify FSRQs and
BL~Lac objects. 

The probability of the BL~Lac type ($P_{\rm BL}$) for each sample
obtained with the classifier using the four parameters is listed in
table~\ref{tab:log} in the Appendix. We can determine the class of
each sample based on $P_{\rm BL}$. Table~\ref{tab:err} is the error
matrix for several different decision criteria: $P_{\rm BL}=0.5$,
$0.6$, and $0.7$. In the case of $P_{\rm BL}=0.5$, all the samples
classified as BL~Lac objects are indeed BL~Lac objects (Accuracy
$=1.0$). On the other hand, only six of the 12 FSRQs are correctly
classified as FSRQ, while the other six samples are misidentified. The
BL~Lac prediction accuracy improves with increased decision criterion
($P_{\rm BL}$), while the prediction accuracy of FSRQs decreases in
that case. The high rate of misidentified FSRQs suggests that a
significant portion of FSRQs cannot be distinguished from BL~Lac
objects based on the four features. 

\begin{table}
  \tbl{Error matrix and accuracy.}{%
    \begin{tabular}{crrr}
      \hline
      \multicolumn{4}{c}{$P_{\rm BL}=0.5$}\\
      \hline
             & \multicolumn{2}{c}{Reference} & Accuracy\\
      Classification & BL Lac & FSRQ & \\
      BL Lac         & 26     & 6    & 0.81\\
      FSRQ           & 0      & 6    & 1.00\\
      \hline
      \multicolumn{4}{c}{$P_{\rm BL}=0.6$}\\
      \hline
             & \multicolumn{2}{c}{Reference} & Accuracy\\
      Classification & BL Lac & FSRQ & \\
      BL Lac         & 25     & 4    & 0.86\\
      FSRQ           & 1      & 8    & 0.89\\
      \hline\multicolumn{4}{c}{$P_{\rm BL}=0.7$}\\
      \hline
             & \multicolumn{2}{c}{Reference} & Accuracy\\
      Classification & BL Lac & FSRQ & \\
      BL Lac         & 21     & 3    & 0.88\\
      FSRQ           & 5      & 9    & 0.64\\
      \hline
\end{tabular}}\label{tab:err}
\end{table}

\section{Discussion}
\subsection{Significance of the classifier}

\begin{figure}
 \begin{center}
  \includegraphics[width=8cm]{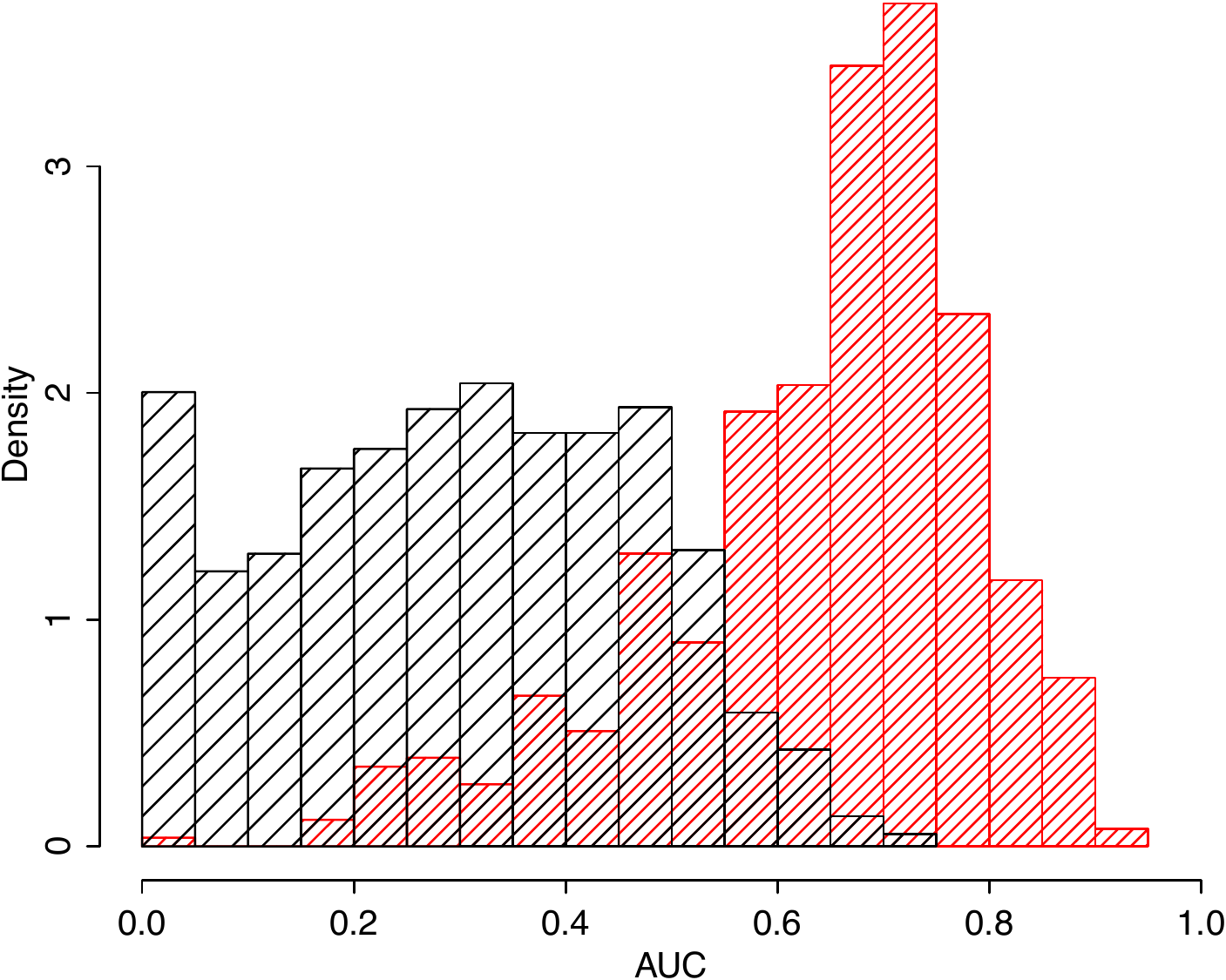} 
 \end{center}
 \caption{Histograms of the AUC obtained from real samples (red)
   and artificial data generated from random numbers (black). See the
   text for details.}\label{fig:auchist}
\end{figure}

A good classifier could incidentally be obtained in high dimensional
problems even if all the features are not related to the real 
characteristics of the samples. We tested the significance of the
obtained classifier described in the previous section using artificial
data sets. The sets of artificial data consist of 12 FSRQs and 26
BL~Lac objects, as for the case in table~\ref{tab:log}, with
random values for the nine features. The random numbers were uniformly
distributed between $0$ and $1$. We made 100 sets of data and obtained
$511\times 100=51100$ AUC values, in the same manner as described in
the previous section. Figure~\ref{fig:auchist} shows histograms of the
AUC values from the real samples (red) and from the artificial samples
(black). The distribution from the real samples exhibited
systematically higher AUCs than that of the random data sets. The AUC
values obtained from the random data sets are concentrated in the area
AUC $<0.8$. Thus, it is unlikely that the obtained best AUC values
from the real data ($\sim 0.9$) are incidentally obtained.

\subsection{Implications from the four features}

\begin{figure}
 \begin{center}
  \includegraphics[width=8cm]{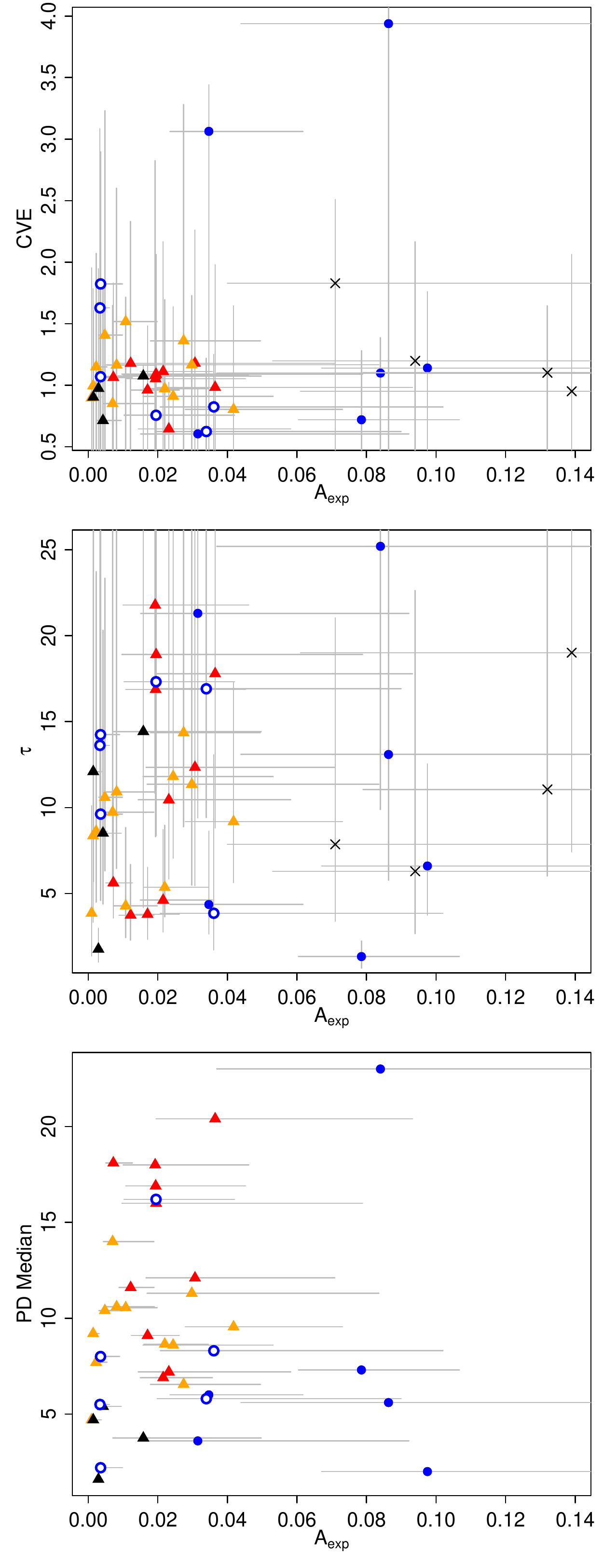} 
 \end{center}
 \caption{Scatter plots of the selected features. The top, middle,
   and bottom panels show CVE, $\tau$, and PD median against
   $A_{\rm exp}$, respectively. The blue filled and open circles
   denote the correctly classified and misclassified FSRQ  samples,
   respectively. The red, orange, and black triangles are LBL, IBL,
   and HBL samples, respectively. The gray bars represent the
   95\% confidence intervals.}\label{fig:scat}
\end{figure}

Here, we discuss the implications of the results of \S~3.
Figure~\ref{fig:scat} shows scatter plots of the four features. The
correctly classified FSRQs ($P_{\rm BL}<0.5$) are indicated by the
filled blue circles, while the misclassified FSRQs are indicated by
the open circles. As can be seen from the top panel, the correctly
classified FSRQs have high CVE and/or high $A_{\rm exp}$, while the
misclassified FSRQs have values for these features comparable to those
of the BL Lac objects. The high value of CVE indicates the presence of prominent
non-stationary flares which deviate from the stationary OU
process. The high value of $A_{\rm exp}$ indicates a large amplitude
of variation at the characteristic timescale, $\tau$. We propose that
FSRQs are characterized by rare and large flares which have a 
time-series structure distinct from ordinary variations. If the
frequency of the flares is relatively high, a few times a year say,
then the light curve can be reproduced by the OU process with a high
$A_{\rm exp}$. If the frequency is low, such as once a year, then the
light curve can be divided into two distinct periods, that is, the
stationary state and the non-stationary flare, which causes a high
CVE. The misclassified FSRQs may be objects in which the flare
frequency was so low that no flare was detected in the year. 

It is not evident that the characteristics of the light-curve CVE
and $A_{\rm exp}$ originate from a different structure and/or
physical condition for the jets between the blazar subtypes
(e.g. \cite{ito16fermi}). It is possible that it is simply due to the
$\nu_{\rm peak}$ effect. In order to investigate this point, we
analyzed the X-ray data of the HBL Mrk~421 using the X-ray light curve
presented in \citet{yam19mrk421}. The data was obtained with
XRT/{\it Swift} from 2009 to 2014. The time-interval of the X-ray light
curve is 1~d. We analyzed the data in the same manner as for the
optical light curves, that is, dividing it into one-year-segments,
calculating the flux density in a logarithmic scale whose linear
trends were subtracted, and performing the OU process regression for
each segment. Table~\ref{tab:xrt} shows the estimated CVE,
$A_{\rm exp}$, and $\tau$ for each sample. We successfully obtained
values for the four segments listed in the table. We could not obtain
those values for the segment MJD~55939--56078 mainly because of the small
sample size ($N=34$). The estimated values are indicated by crosses in
the top and middle panels of figure~\ref{fig:scat}. They are
definitely in the regime of FSRQs, especially regarding the large
$A_{\rm exp}$. This result suggests that the large $A_{\rm exp}$ does
not originate from different jet properties in the blazar subtypes,
but from the $\nu_{\rm peak}$ effect.

The variation timescale of the light-curve, $\tau$, was also selected
as an important feature for classification. However, as shown in the
middle panel of figure~\ref{fig:scat}, we cannot find any clear
differences between the $\tau$ distributions of FSRQs and BL Lac
objects. This feature was selected mainly because it is useful for the
classification of only one FSRQ sample, 3C~454.3 in
MJD~54542--54930. This sample has a small CVE ($=0.60$) and a not very
high $A_{\rm exp}$ ($=3.15$), from which the object cannot be
distinguished from BL Lac objects, but has an exceptionally large
$\tau$ ($=21.30$). We consider that our analysis does not provide
enough evidence to determine the importance of $\tau$.

It is proposed that the beaming factor of FSRQs is systematically
larger than that of BL~Lac objects. \citet{gio12type} proposed that
the two subtypes have a common nature, except for the beaming
factor. \citet{ito16fermi} propose that the jet volume fraction
occupied by the fast ``spine'' should be larger in FSRQs than BL~Lac
objects. The difference in the beaming factor would also change the
characteristics of the variability. For example, a shorter variation
timescale is expected with a higher beaming factor. However, our
analysis provides no strong evidence that $\tau$ of FSRQs is
systematically smaller than that of BL~Lac objects, while the
uncertainty of $\tau$ is large. 

\begin{figure}
 \begin{center}
  \includegraphics[width=8cm]{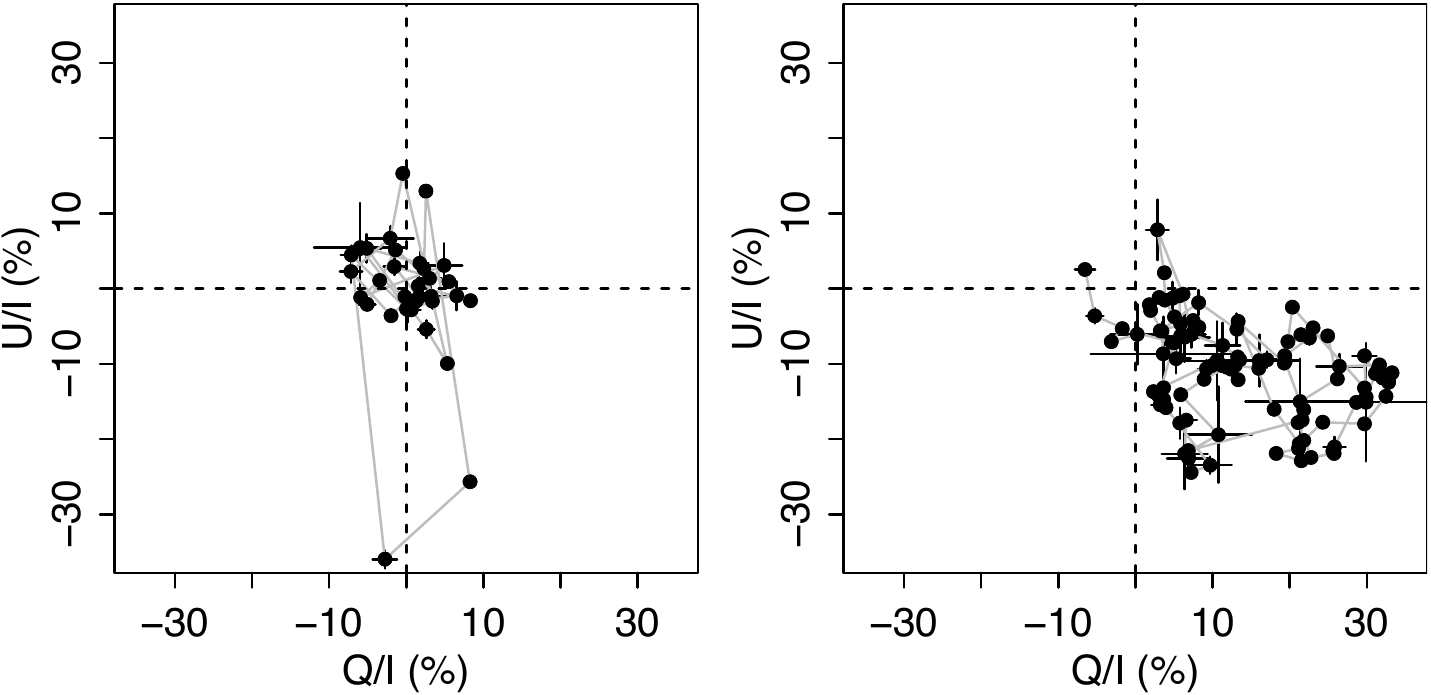} 
 \end{center}
 \caption{Polarization variations in the Stokes $Q/I$--$U/I$ plane of
   the FSRQ PKS~1510$-$089 in MJD~54759--55124 (left) and the LBL
   OJ~287 in MJD~55045--55451 (right).}\label{fig:qusample} 
\end{figure}

In the bottom panel of figure~\ref{fig:scat}, we can see a trend that
BL Lac objects have high PD medians compared with FSRQs. This
characteristic is stronger when HBLs (the black triangles in the
figure) are neglected. The low PD medians of HBLs are probably due to
a large contamination of the unpolarized emission from their host
galaxies (\cite{sha13spec}). Figure~\ref{fig:qusample} shows examples
of polarization variations in the Stokes $Q/I$--$U/I$ plane. The
left and right panels show the data of the FSRQ PKS~1510$-$089 having
a low PD median and that of the LBL OJ~287 having a high PD median,
respectively. An increase in PD is occasionally associated with the
flares of blazars, and in general the PD remains relatively low when the
object is faint. In the left panel of figure~\ref{fig:qusample}, most
of the data points exhibit low PDs, except for a few data points with
high PDs over $>20$\%, which are associated with the prominent flare
shown in panel~(a) of figure~\ref{fig:samples}. The high value of the
PD median in BL~Lac objects indicates that PD is relatively high even
in the faint state. The right panel of figure~\ref{fig:qusample} shows
an example: the object had a relatively high PD throughout the
year. It has been proposed that the LBL object BL~Lac has two
polarization components: short-term variations superimposed on a
stable or semi-stable component (\cite{hag02bllac,sak13bllac}). The
fact that the PD median was selected in our analysis suggests that the
presence of a stable polarization component is a characteristic
feature of BL Lac objects. 

The origin of the PD median characteristic is unclear. The values of
the PD median apparently correlate with $\nu_{\rm peak}$ in BL~Lac
objects, being lowest in HBLs and highest in LBLs. However, FSRQs have
low PD medians, although they have the highest $\nu_{\rm peak}$.
Hence, the characteristic is not due to the $\nu_{\rm peak}$ effect,
but possibly due to a difference in the jet structure between the
blazar subtypes. In this case, the stable polarization component
should have a different emitting site or physical condition from the
short-term variations since the characteristics of the short-term flux
variability can be interpreted as the $\nu_{\rm peak}$ effect, as
discussed above. The presence of the stable component may suggest that
the accelerated electrons have a long life-time with a long cooling
timescale. According to \citet{kas05blr}, the size of the broad line 
region (BLR) has a positive correlation with the AGN luminosity, and
the AGNs with the highest luminosity have large BLRs up to sub-pc
scale. FSRQs form a sub-group of blazars with the highest luminosity,
not only of the jets, but also of the AGNs
(\cite{fos98blseq,sha13spec,ghi17blseq}). The lack of a stable
component in FSRQs may be reconciled with the presence of a strong
radiation field induced by a large BLR causing strong Compton cooling
of the electrons even in the sub-pc region, which is the source of the
stable polarization component in BL~Lac objects. 

\citet{gio12type} reported that LBLs include both low luminosity FR~I
objects and high luminosity FR~II objects. If this is the case, there
may be LBLs with a low PD median. Our samples included only two LBL
objects, BL~Lac and OJ~287. The number of samples is so small
that we cannot make conclusions about the population of LBLs. Further
studies are required to understand the relationship between the
presence/absence of the stable polarization component and FR types
or AGN luminosity. 
  
\subsection{Features of polarization variability}

\begin{figure}
 \begin{center}
  \includegraphics[width=8cm]{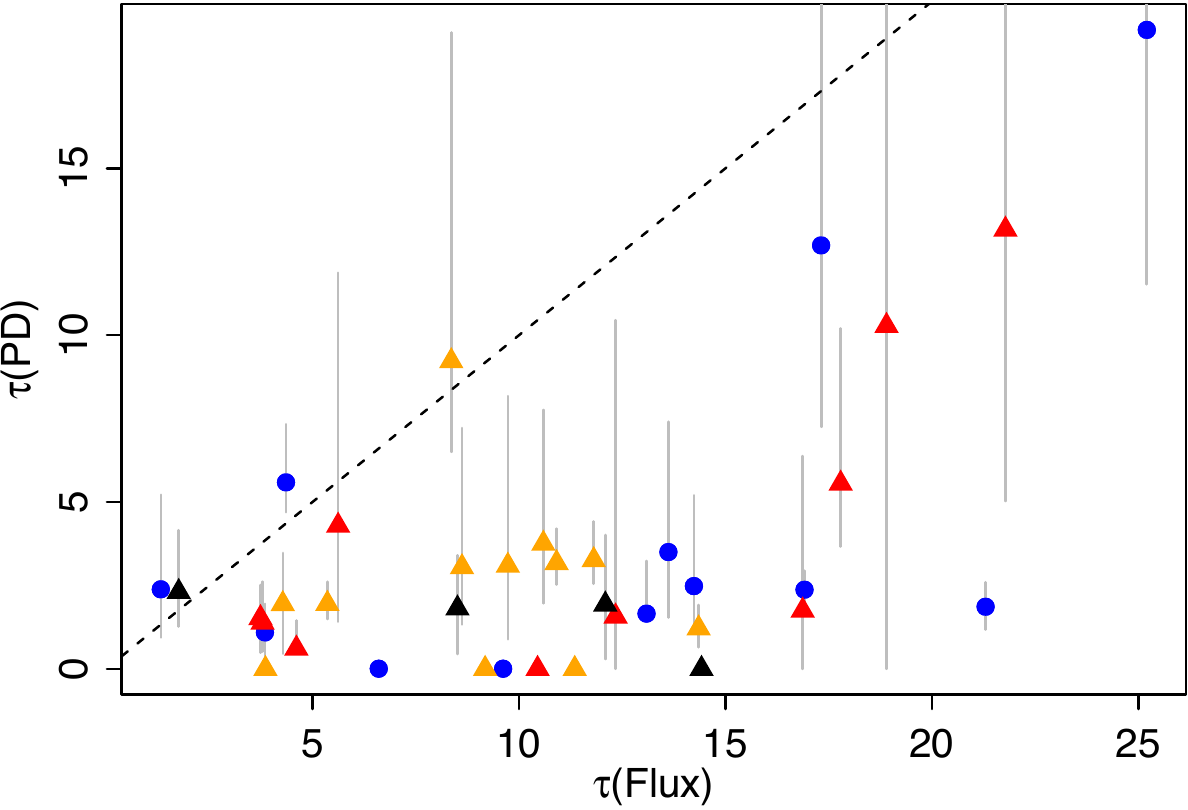} 
 \end{center}
 \caption{Correlations of the flux and PD $\tau$. The blue
   circles and the red, orange, and black triangles denote the FSRQ, LBL,
   IBL, and HBL samples, respectively. The dashed line indicates
   $\tau\;{\rm (PD)}=\tau\;{\rm (Flux)}$.}\label{fig:flxpd} 
\end{figure}

In this paper, we used features derived from both the light curves and
PD variations, while the only PD feature selected as being useful
for classifying FSRQs and BL~Lac objects was the PD median.
Figure~\ref{fig:flxpd} shows a scatter plot of $\tau$ of the light
curve and of the PD variations. In this figure, we can see that the
timescale of the PD variation tends to be shorter than that of the
light curve. Most of the objects have a PD $\tau$ shorter than 5~d. As
mentioned in \S~2.2, the PD $\tau$ was too short to be uniquely
determined in seven cases. These results imply that the real $\tau$
could be too short to be correctly estimated from our data. If this is
the case, the PD features were not selected in our analysis possibly
because they were not good indicators for the nature of the PD
variability. The fact that the PD $\tau$ is significantly shorter than
the light-curve $\tau$ suggests that the relaxation timescale of the
ordered magnetic field is shorter than the cooling timescale of the
accelerated electrons. 

On the other hand, the presence of a stable polarization component in
BL~Lac objects can also cause a lack of PD variation features in the
selected features. The observed Stokes parameters are a sum of those
of multiple components. If the contamination of the stable component
is strong, the PD variation of a flare is diluted by the stable
component. For example, the increase in PD associated with a flare is
canceled if the direction of polarization of the flare component is
perpendicular to that of the stable component. This effect also causes
the PD features  to be poor indicators of the real PD variability. In
future work, we will extract the features of the PD variation for both
the short-term flares and the stable, or long-term, variation
component by separating these components (\cite{uem10bayes}). 

\section{Summary}
We characterized the optical variability of blazars using the OU
process and investigated the features which are discriminative for the
two blazar subtypes, FSRQs and BL~Lac objects. Our summarized findings
are as follows:  
\begin{itemize}
\item Four features, namely, the variation amplitude, 
  $A_{\rm exp}$, characteristic timescale, $\tau$, and
  non-stationarity, CVE, from the light curve and the PD median 
  are essential to classify blazars into FSRQs and BL~Lac objects.
\item FSRQs are characterized by rare and large flares based on a
  large $A_{\rm exp}$ and/or CVE. We found that the X-ray variability
  of the HBL Mrk~421 also has large $A_{\rm exp}$ comparable to the
  optical variability of FSRQs. Hence, the characteristics of
  $A_{\rm exp}$ and CVE are governed not by the differences in the jet
  structure between the subtypes, but by the $\nu_{\rm peak}$ effect.
\item The high PD median of BL~Lac objects suggest that they tend to
  have a stable polarization component. The lack of such a component
  in FSRQs is possibly due to strong Compton cooling from a
  large BLR in sub-pc scale jets.
\item The variation timescale of PD is significantly shorter than
  that of the light curves. This may indicate that the relaxation
  timescale of the ordered magnetic field is shorter than the cooling
  timescale of the accelerated electrons.
\end{itemize}

%%%%%%%%%%%%%%%%%%%%%%%%%%%%%%%%%%%%%%%

\begin{ack}
This work was supported by a Kakenhi Grant-in-Aid (No. 25120007) from
the Japan Society for the Promotion of Science (JSPS).  
\end{ack}

\appendix 
%\section*{Case of single paragraph}
\section{Samples and features}
\begin{landscape}
\tabcolsep = 1mm
\begin{table}
  % P_BL : smlr_tmp25/smlr.summary.test1.txt
  \tbl{Samples and features for classification}{%
    %%% for two columns
    %\footnotesize
    %%% for draft mode
    \scriptsize
    \begin{tabular}{crrrrrrrrrrrrrcr}
    \hline
    Object & Start & End & N & \multicolumn{4}{c}{Light curve} & &
    \multicolumn{5}{c}{Polarization degree} & Type & $P_{\rm BL}$\\
    \cline{2-3} \cline{5-8} \cline{10-14}
    & \multicolumn{2}{c}{MJD} &  & CVE &
    Slope & $A_{\rm exp}$ & $\tau$ & & Median & CVE &
    Slope & $A_{\rm exp}$ & $\tau$ & &\\
    & & & & & ($\times 10^{-4}$) & ($\times 10^{-2}$) & & & & &
    ($\times 10^{-3}$) & ($\times 10^{-2}$) & & \\
    \hline
    S2 0109$+$22 & 54581 & 54974 &  69 & $1.52_{-1.47}^{+0.20}$ &
    $6.00\pm 2.23$ & $1.08_{-0.36}^{+0.91}$ & $4.28_{-1.85}^{+4.56}$ &
    &10.55 & $1.18_{-1.11}^{+0.57}$ & $-2.08\pm 0.65$ &
    $7.12_{-2.00}^{+3.31}$ & $1.95_{-1.49}^{+1.53}$ & IBL & 0.86\\
    3C 66A       & 54612 & 54974 & 114 & $1.16_{-1.12}^{+1.44}$ &
    $-0.86\pm 1.11$ & $0.82_{-0.30}^{+1.09}$ & $10.91_{-4.46}^{+16.11}$&
    &10.60 & $1.24_{-1.17}^{+0.25}$ & $-2.72\pm 0.26$ &
    $6.63_{-0.79}^{+1.21}$ & $3.17_{-0.65}^{+1.03}$ & IBL & 0.86\\
    3C 66A       & 54974 & 55345 & 142 & $1.41_{-1.27}^{+1.82}$ &
    $9.57\pm 0.66$ & $0.48_{-0.17}^{+0.51}$ & $10.59_{-4.28}^{+12.76}$&
    &10.40 & $1.02_{-0.94}^{+0.40}$ & $-0.86\pm 0.23$ &
    $6.71_{-1.05}^{+2.40}$ & $3.75_{-1.79}^{+4.01}$ & IBL & 0.85\\
    3C 66A       & 55345 & 55671 &  72 & $0.85_{-0.81}^{+0.80}$ &
    $-2.03\pm 1.28$ & $0.70_{-0.27}^{+1.19}$ &$9.73_{-4.31}^{+17.90}$&
    &14.00 & $1.31_{-1.19}^{+1.17}$ & $1.03\pm 0.24$ &
    $6.54_{-0.83}^{+2.02}$ & $3.10_{-2.21}^{+5.08}$ & IBL & 0.90\\
    3C 66A       & 55671 & 56073 &  30 & $1.00_{-0.90}^{+0.79}$ &
    $3.65\pm 1.17$ & $0.14_{-0.06}^{+0.18}$ & $8.36_{-5.04}^{+12.82}$&
    & 9.20 & $1.32_{-1.26}^{+0.63}$ & $-4.66\pm 1.55$ &
    $7.21_{-4.56}^{+19.20}$ & $9.22_{-2.72}^{+9.85}$ & IBL & 0.85\\
    3C 66A       & 56409 & 56797 &  56 & $1.15_{-0.80}^{+0.92}$ &
    $-4.31\pm 1.06$ & $0.23_{-0.09}^{+0.32}$ &$8.62_{-4.13}^{+15.11}$&
    &7.70 & $1.19_{-1.06}^{+1.17}$ & $-1.45\pm 0.25$ &
    $5.95_{-0.35}^{+0.86}$ & $3.05_{-1.72}^{+4.17}$ & IBL & 0.81\\
    S5 0716$+$714& 54665 & 55050 & 138 & $1.36_{-1.32}^{+1.92}$ &
    $-8.45\pm 1.57$ & $2.74_{-0.96}^{+2.21}$&$14.35_{-5.47}^{+11.99}$&
    &6.55 & $1.14_{-0.96}^{+0.85}$ & $-0.11\pm 0.28$ &
    $6.66_{-1.53}^{+2.24}$ & $1.22_{-0.57}^{+0.70}$ & IBL & 0.68\\
    S5 0716$+$714& 55050 & 55389 & 156 & $0.97_{-0.84}^{+0.72}$ &
    $-7.20\pm 1.39$ & $2.20_{-0.61}^{+1.26}$ &$5.36_{-1.72}^{+3.61}$ &
    &8.65 & $1.15_{-0.93}^{+0.67}$ & $0.57\pm 0.32$ &
    $7.07_{-1.80}^{+2.60}$ & $1.95_{-0.46}^{+0.66}$ & IBL & 0.83\\
    S5 0716$+$714& 55389 & 55746 &  98 & $0.91_{-0.76}^{+0.73}$ &
    $-17.70\pm 1.84$ & $2.44_{-0.87}^{+2.88}$&$11.81_{-4.76}^{+15.96}$&
    &8.60 & $1.18_{-1.04}^{+0.80}$ & $0.54\pm 0.38$ &
    $7.00_{-1.79}^{+2.98}$ & $3.26_{-0.71}^{+1.16}$ & IBL & 0.79\\
    S5 0716$+$714& 55746 & 56125 &  98 & $0.80_{-0.76}^{+0.84}$ &
    $10.99\pm 2.16$ & $4.18_{-1.39}^{+3.13}$ & $9.18_{-3.55}^{+8.14}$&
    &9.55 & $1.18_{-1.11}^{+0.21}$ & $-2.09\pm 0.60$ &
    $6.82_{-1.69}^{+2.51}$ & $0.00$ & IBL & 0.78\\
    S5 0716$+$714& 56484 & 56868 &  51 & $1.17_{-1.10}^{+0.56}$ &
    $16.45\pm 3.37$ & $2.98_{-1.29}^{+5.38}$&$11.35_{-5.87}^{+22.98}$&
    &11.30 & $1.14_{-1.12}^{+0.56}$ & $0.32\pm 0.52$ &
    $7.03_{-2.02}^{+3.26}$ & $0.00$ & IBL & 0.84\\
    OJ 287       & 54696 & 55045 & 100 & $0.98_{-0.84}^{+1.00}$ &
    $-12.39\pm 1.77$ &$3.65_{-1.70}^{+5.68}$&$17.79_{-8.98}^{+28.97}$&
    &20.40 & $1.16_{-1.06}^{+0.68}$ & $1.89\pm 0.18$ &
    $7.30_{-0.44}^{+1.10}$ & $5.55_{-1.89}^{+4.65}$ & LBL & 0.76\\
    OJ 287       & 55045 & 55451 & 116 & $1.08_{-1.04}^{+1.75}$ &
    $-16.84\pm 1.58$ &$1.92_{-0.92}^{2.69}$&$21.78_{-10.96}^{+31.06}$&
    &18.00 & $1.13_{-1.10}^{+0.65}$ & $-4.98\pm 0.26$ &
    $7.38_{-1.93}^{+8.49}$ & $13.17_{-8.14}^{+33.42}$ & LBL & 0.69\\
    OJ 287       & 55773 & 56161 &  42 & $1.09_{-1.06}^{+0.97}$ &
    $14.54\pm 2.26$ &$1.95_{-0.99}^{+5.94}$&$18.90_{-10.54}^{+63.15}$&
    &16.00 & $1.35_{-1.23}^{+1.48}$ & $2.64\pm 0.75$ &
    $7.18_{-3.21}^{+10.63}$ & $10.28_{-10.28}^{+41.50}$ & LBL & 0.77\\
    OJ 287       & 56488 & 56872 &  54 & $1.05_{-1.01}^{+0.88}$ &
    $5.73\pm 2.77$ & $1.94_{-0.86}^{+2.59}$ &$16.87_{-8.57}^{+24.93}$&
    &16.90 & $1.27_{-1.22}^{+0.14}$ & $0.47\pm 0.33$ &
    $6.92_{-0.68}^{+1.52}$ & $1.75_{-1.75}^{+4.63}$ & LBL & 0.83\\
    Mrk 421      & 55127 & 55438 &  36 & $1.08_{-0.97}^{+0.08}$ &
    $-23.73\pm 2.76$ & $1.58_{-0.88}^{+3.39}$&$14.42_{-10.24}^{+32.96}$&
    &3.75 & $1.16_{-1.14}^{+0.30}$ & $-0.49\pm 0.78$ &
    $5.48_{-1.89}^{+6.63}$ & $0.00$ & HBL & 0.61\\
    3C 279       & 54621 & 55115 &  58 & $0.76_{-0.70}^{+0.86}$ &
    $-38.92\pm 2.62$ & $1.95_{-0.92}^{+2.26}$&$17.32_{-8.94}^{+21.41}$&
    &16.20 & $1.06_{-0.82}^{+0.31}$ & $-5.09\pm 0.65$ &
    $7.69_{-3.31}^{+16.86}$ & $12.69_{-5.44}^{+27.37}$ & FSRQ & 0.81\\
    PKS 1502$+$106& 54568 & 55045 &  60 & $1.10_{-0.89}^{+0.29}$ &
    $6.85\pm 2.19$ &$8.40_{-4.71}^{+26.42}$&$25.20_{-15.32}^{+87.56}$&
    &23.00 & $1.05_{-0.88}^{+0.29}$ & $1.76\pm 0.32$ &
    $8.01_{-4.74}^{+17.97}$ & $19.15_{-7.63}^{+28.63}$ & FSRQ & 0.28\\
    PKS 1510$-$089& 54759 & 55124 &  50 & $3.94_{-3.89}^{+1.96}$ &
    $3.74\pm 8.65$ & $8.63_{-4.25}^{+16.53}$&$13.09_{-7.32}^{+27.56}$&
    &5.60 & $1.61_{-1.20}^{+0.64}$ & $-0.03\pm 1.15$ &
    $7.22_{-2.44}^{+4.01}$ & $1.66_{-0.19}^{+1.58}$ & FSRQ & 0.12\\
    RX J1542.8$+$612& 54895 & 55490 & 99 & $1.63_{-1.56}^{+1.46}$ &
    $-0.63\pm 0.42$ & $0.34_{-0.11}^{+0.28}$&$13.62_{-5.07}^{+12.45}$&
    &5.50 & $1.09_{-0.96}^{+0.40}$ & $-0.28\pm 0.22$ &
    $6.38_{-2.87}^{+5.46}$ & $3.50_{-1.96}^{+3.91}$ & FSRQ & 0.66\\
    PG 1553$+$113& 56629 & 56956 &  50 & $0.90_{-0.70}^{+0.70}$ &
    $2.51\pm 0.77$ & $0.14_{-0.06}^{+0.24}$ &$12.09_{-6.41}^{+22.26}$&
    &4.70 & $1.21_{-1.08}^{+0.73}$ & $-0.47\pm 0.58$ &
    $5.87_{-1.57}^{+3.00}$ & $1.93_{-1.64}^{+2.08}$ & HBL & 0.68\\
    Mrk 501      & 56289 & 56660 &  84 & $0.97_{-0.94}^{+0.97}$ &
    $-1.73 \pm 0.76$ & $0.29_{-0.08}^{+0.12}$ &$1.76_{-0.76}^{+1.22}$&
    &1.60 & $1.22_{-1.03}^{+1.54}$ & $0.03\pm 0.27$ &
    $5.14_{-1.18}^{+2.25}$ & $2.31_{-1.04}^{+1.85}$ & HBL & 0.59\\
    PKS 1749$+$096& 55880 & 56780 & 77 & $1.14_{-0.78}^{+0.62}$ &
    $8.38\pm 2.04$ & $9.75_{-3.04}^{+6.14}$ &$6.60_{-2.86}^{+5.94}$&
    &2.00 & $1.35_{-1.25}^{+2.12}$ & $-0.11\pm 1.06$ &
    $5.64_{-0.05}^{+0.12}$ & $0.00$ & FSRQ & 0.25\\
    3C 371       & 54555 & 54952 &  81 & $1.07_{-0.95}^{+1.29}$ &
    $2.78 \pm 1.00$ & $0.35_{-0.16}^{+0.55}$&$14.24_{-6.53}^{+20.37}$&
    &8.00 & $1.14_{-0.98}^{+1.01}$ & $-0.10\pm 0.19$ &
    $5.90_{-0.27}^{+0.52}$ & $2.48_{-1.38}^{+2.72}$ & FSRQ & 0.76\\
    1ES 1959$+$650& 54577 & 54934 & 40 & $0.90_{-0.86}^{+1.06}$ &
    $5.76 \pm 0.94$ & $0.10_{-0.04}^{+0.10}$ & $3.86_{-2.50}^{+6.25}$&
    &4.70 & $1.16_{-1.07}^{+0.71}$ & $-0.26\pm 0.55$ &
    $5.73_{-0.71}^{+1.21}$ & $0.00$ & IBL & 0.75\\
    PKS 2155$-$304& 54628 & 54925 & 57 & $0.71_{-0.67}^{+1.08}$ &
    $23.68\pm 1.75$ & $0.42_{-0.18}^{+0.53}$ &$8.51_{-4.13}^{+11.81}$&
    &5.40 & $1.24_{-1.22}^{+0.31}$ & $0.10\pm 0.55$ &
    $5.92_{-0.93}^{+1.81}$ & $1.81_{-1.37}^{+1.59}$ & HBL & 0.74\\
    BL Lac        & 54537 & 54908 & 100 & $1.06_{-0.98}^{+0.77}$ &
    $-1.75\pm 1.31$ & $0.72_{-0.22}^{+0.55}$ &$5.62_{-2.05}^{+5.15}$&
    &18.10 & $1.03_{-0.73}^{+0.88}$ & $-0.09\pm 0.24$ &
    $7.16_{-0.96}^{+2.61}$ & $4.29_{-2.88}^{+7.58}$ & LBL & 0.92\\
    BL Lac        & 54908 & 55270 & 104 & $1.18_{-0.99}^{+1.15}$ &
    $8.11\pm 1.23$ & $1.22_{-0.34}^{+0.68}$ &$3.74_{-1.46}^{+2.95}$&
    &11.60 & $1.14_{-1.04}^{+0.61}$ & $-0.53\pm 0.22$ &
    $6.84_{-0.70}^{+1.01}$ & $1.52_{-1.04}^{+0.99}$ & LBL & 0.89\\
    BL Lac        & 55270 & 55605 &  58 & $1.18_{-0.83}^{+1.08}$ &
    $-13.33\pm 2.30$ & $3.07_{-1.41}^{+4.02}$&$12.34_{-6.89}^{+17.89}$&
    &12.10 & $1.19_{-0.57}^{+0.44}$ & $1.49\pm 0.47$ &
    $6.84_{-3.11}^{+10.31}$ & $1.57_{-1.57}^{+8.88}$ & LBL & 0.84\\
    BL Lac        & 55605 & 56011 &  67 & $0.64_{-0.57}^{+0.38}$ &
    $0.85\pm 2.28$ & $2.32_{-0.88}^{+3.51}$ &$10.45_{-4.61}^{+16.48}$&
    &7.20 & $1.14_{-0.98}^{+0.58}$ & $0.17\pm 0.48$ &
    $6.71_{-1.40}^{+2.05}$ & $0.00$ & LBL & 0.76\\
    BL Lac        & 56011 & 56347 &  82 & $1.11_{-1.05}^{+1.05}$ &
    $-26.64\pm 2.01$ & $2.15_{-0.66}^{+1.42}$ &$4.61_{-1.85}^{+4.11}$&
    &6.90 & $1.20_{-1.18}^{+0.27}$ & $-2.22\pm 1.08$ &
    $6.69_{-3.65}^{+5.49}$ & $0.61_{-0.07}^{+0.84}$ & LBL & 0.79\\
    BL Lac        & 56347 & 56735 & 100 & $0.96_{-0.89}^{+0.52}$ &
    $21.72\pm 1.45$ & $1.70_{-0.47}^{+0.92}$ &$3.79_{-1.46}^{+2.73}$&
    &9.10 & $1.19_{-1.01}^{+0.52}$ & $0.63\pm 0.35$ &
    $6.86_{-1.76}^{+2.80}$ & $1.39_{-0.87}^{+1.23}$ & LBL & 0.85\\
    CTA 102       & 56140 & 56339 &  41 & $0.82_{-0.78}^{+0.43}$ &
    $-40.03\pm 9.38$ & $3.61_{-1.54}^{+6.59}$ &$3.85_{-2.14}^{+9.22}$&
    &8.30 & $1.87_{-1.80}^{+2.25}$ & $7.44\pm 2.09$ &
    $7.18_{-1.92}^{+3.17}$ & $1.09_{-1.00}^{+0.86}$ & FSRQ & 0.79\\
    3C 454.3      & 54542 & 54930 & 108 & $0.60_{-0.55}^{+0.36}$ &
    $-39.57\pm 1.84$&$3.15_{-1.64}^{+6.07}$&$21.30_{-11.91}^{+42.74}$&
    &3.60 & $1.29_{-1.28}^{+0.55}$ & $-1.01\pm 0.51$ &
    $6.41_{-1.90}^{+2.83}$ & $1.87_{-0.69}^{+0.73}$ & FSRQ & 0.39\\
    3C 454.3      & 54930 & 55270 & 155 & $0.62_{-0.59}^{+0.60}$ &
    $16.01\pm 1.65$ & $3.39_{-1.41}^{+5.60}$&$16.91_{-7.50}^{+30.43}$&
    &5.80 & $1.32_{-1.06}^{+0.48}$ & $0.91\pm 0.34$ &
    $6.73_{-1.70}^{+2.48}$ & $2.37_{-0.42}^{+0.57}$ & FSRQ & 0.57\\
    3C 454.3      & 55270 & 55795 &  91 & $0.72_{-0.93}^{+0.56}$ &
    $0.72\pm 2.94$ & $7.85_{-1.81}^{+2.82}$ &$1.33_{-0.68}^{+0.91}$&
    &7.30 & $1.42_{-1.00}^{+0.91}$ & $0.45\pm 0.32$ &
    $7.23_{-2.68}^{+5.18}$ & $2.38_{-1.44}^{+2.84}$ & FSRQ & 0.48\\
    3C 454.3      & 56000 & 56370 &  53 & $1.82_{-1.79}^{+1.07}$ &
    $10.97\pm 1.34$ & $0.36_{-0.16}^{+0.63}$ &$9.62_{-5.03}^{+19.83}$&
    &2.20 & $1.19_{-1.16}^{+0.38}$ & $-0.82\pm 4.32$ &
    $5.81_{-4.13}^{+19.51}$ & $0.00$ & FSRQ & 0.56\\
    3C 454.3      & 56370 & 56744 &  78 & $3.06_{-3.26}^{+0.38}$ &
    $4.09\pm 3.81$ & $3.47_{-1.12}^{+2.71}$ &$4.36_{-1.72}^{+4.27}$&
    &6.00 & $1.61_{-1.24}^{+1.97}$ & $0.48\pm 0.83$ &
    $7.21_{-4.75}^{+8.85}$ & $5.59_{-0.90}^{+1.74}$ & FSRQ & 0.37\\
      \hline
\end{tabular}}\label{tab:log}
\end{table}
\end{landscape}

\begin{table}
  % result_xrt.txt
  \tbl{Samples and features of Mrk~421.}{%
    \begin{tabular}{rrrrrr}
      \hline
      \multicolumn{2}{c}{MJD} & N & CVE & $A_{\rm exp}$ & $\tau$ \\
      \hline
      55150 & 55386 & 119 & $1.10_{-1.06}^{+0.54}$ &
      $0.13_{-0.05}^{+0.20}$ & $11.05_{-5.04}^{+20.53}$ \\
      55533 & 55651 & 58  & $0.95_{-0.89}^{+1.11}$ &
      $0.14_{-0.08}^{+0.30}$ & $19.01_{-11.59}^{+38.48}$\\
      56268 & 56428 & 54  & $1.83_{-1.79}^{+0.68}$ &
      $0.07_{-0.03}^{+0.08}$ & $7.86_{-4.48}^{+13.18}$ \\
      56627 & 56749 & 42  & $1.20_{-1.15}^{+0.97}$ &
      $0.09_{-0.04}^{+0.18}$ & $6.29_{-3.63}^{+16.34}$ \\
      \hline
\end{tabular}}\label{tab:xrt}
\end{table}

%\section{Case of two or more paragraphs}

%%%
% See the manual for the detail.
%%%
%\bibliographystyle{pasjtest1}
%\bibliography{blazar}

\end{document}